
\magnification=\magstep1
\baselineskip  22 true   pt
\hsize 5 in\hoffset=.4 true in
\vsize 6.9 in\voffset=.4  true  in

\def\a{\alpha}

\def\d{\delta}
\def\e{\epsilon}

\def\l{\lambda}
\def\om{\omega}
\def\p{\pi}
\def\m{\mu}
\def\n{\nu}
\def\r{\rho}

\def\D{$R^3 X S^1$}

\def\G{$\Omega (\vec {x})$}

\def\J{$N_f$}
\def\K{$1\over{g^2R}$}

\def\M{$\{\l (\vec x)\}$}
\def\N{$\{\l^{\infty}\}$}
\def\O{$H_{\{\lambda^{\infty}\}}$}
\def\P{$U(1)^k {\bigotimes}^k_{\a=0} SU(m_{\a})$}
\def\Q{$m_{\a}$}
\def\R{$\mu_{\a}$}
\def\S{($\a = 0,1 \ldots k)$}
\def\T{$r\longrightarrow \infty$}

\def\BB{$V(\vec x)$}
\def\DD{$\omega(\vec x)$}

\def\GG{$\cal T$}

\def\II{${{\bar\eta}_{a\mu\n}}$}

\def\VV{$(x_0,x_1,x_2,x_3)$}

\def\XX{$(\mu=1,2,5,6)$}
\def\YY{$R^2 X S^1 X S^1$}
\def\ZZ{$S_{\infty}^1 X S^1 X S^1$}

\def\QNA{\phi_i(x,y)}
\def\QNAB{\{\phi_i(x,y)\}}
\def\QNB{f^i_n(y)}

\def\QNE{{F^a_{\mu\n}}}
\def\QNEB{{\tilde{F^a_{\mu\n}}}}
\def\QNG{\Omega (\vec {x})}
\def\QNH{q_{\a}}

\def\QNK{{1\over{g^2R}}}

\def\QNR{\mu_{\a}}

\def\QNU{\p_2(G/H)}
\def\QNV{\p_2(H)}
\def\QNX{{\cal Z}^k}

\def\QNBB{V(\vec x)}
\def\QNDD{\omega(\vec x)}
\def\QNEE{d^2{\vec S}}

\def\QNGG{\cal T}

\def\QNII{{{\bar\eta}_{a\mu\n}}}
\def\QNJJ{A_{\m}}
\def\QNKK{{\tilde r}}
\def\QNLL{{\tilde x}^5}

\def\QNOO{{{rR}\over{\p\r^2}}}
\def\QNPP{(\d^{ia} - {{3x^ix^a}\over{r^2}})}
\def\QNQQ{{\rho^2}\over{R}}

\def\QNSS{\psi_i}
\def\QNTT{\psi_i^{\dagger}}

\def\QNBBB{{\tilde{y}_1}}
\def\QNCCC{S^1 X S^1 X S^1 X Z_2}

\def\lnrt{\longrightarrow}

\vfil
\headline{\hfill IP/BBSR/92-14.}
\centerline {\bf{ Syncyclons}}
\centerline {\bf{  or }}
\centerline {\bf{  Solitonic Signals from Extra Dimensions}}
\vfil
\centerline {\bf{C.S.Aulakh}}
\centerline{ Institute of Physics, Sachivalya Marg}
\centerline{ Bhubhaneshwar,751005,India}
\vfil
\centerline{\bf{ ABSTRACT}}

\noindent

   In theories where spacetime is a direct product of Minkowski
  space ($M^4$) and a d dimensional compact space ($K^d$), there can exist
  topological solitons  that  simultaneously wind around
  $R^3$ (or $R^2$ or $R^1$) in $M^4$ and the compact dimensions.
A paradigmatic non-gravitational example of such  ``co-winding"
solitons is furnished by Yang-Mills theory defined on $M^4 X S^1$.
Pointlike, stringlike and sheetlike solitons can be identified by transcribing
and generalizing the proceedure used to construct the periodic
instanton (caloron) solutions. Asymptotically the classical pointlike objects
have non-Abelian magnetic dipole fields together with a  non-Abelian
scalar potential while the ``color" electric charge is zero. However
quantization of collective coordinates associated with zeromodes and
coupling to fermions can radically change these quantum numbers
due to fermion number fractionalization and its non-Abelian generalization.
Interpreting the YM group as color (or the Electroweak SU(2) group)
and assuming that an extra circular dimension exists thus implies the
existence of topologically stable solitonic objects which carry baryon(lepton)
number and a mass O($1/g^2R$), where R is the radius of the compact dimension.

PACS Nos : 12.10. -g,04.20. Jb, 04.60.+n, 14.80.Hv\hfil\break

The speculation that space-time enjoys an extension from Minkowski
space $M^4$ to $M^4 X K^d$ where  $K$ is some d-dimensional compact space is
a common ingredient of the Kaluza-Klein and Superstring scenarios.
However , so far, generic signals of the presence of extra dimensions
are lamentably absent. The reasons for this sorry divorce between
speculation and its experimental verifiability are most easily
explained in terms of field theoretic {\hbox{framework ${}^1$}} developed for
Kaluza-Klein theory. Modulo intrinsically
stringy effects , what follows holds also for the low-energy
behaviour of string theory.

In Kaluza-Klein theory  physics is taken to be governed by an action
which is a functional of fields  ${\phi_i(x,y)}$ defined on $M^4 X
K^d $.
The compact space $K$ is characterized by some radii ${R_1,R_2,..R_d}$.
Physics at energy scales $<<R_i^{-1}$ is extracted by expanding the
fluctuation field around an appropriate  background ${{\bar\phi}_i}$
in harmonics $\{f_n^i(y)\}$ on the compact space $K$ :

$$\QNA = {{\bar{\phi}}_i} + \sum_n{\phi_i^n(x) \QNB}    \eqno(1) $$

The harmonics $\{f_n^i(y)\}$  are eigenfunctions of the appropriate laplacian
on $K$ :

$$ -\bigtriangledown^2_i f^n_i(y) = ``({{n^2}\over {R^2}})"
f^n_i(y)\eqno (2) $$

Here $``n^2"$ is the appropriate ``Casimir" and $R$ the appropriate
combination of length scales associated with $K$. On integrating the
part of the action quadratic in the small fluctuations over
the compact coordinates $\{y^a\}$ one gets a spectrum of small
fluctuations which consists of massless modes $\{\phi_i^{(0)}\}$
($``n^2"=0$) and massive modes $\phi_i^n(x) (``n^2"\neq 0) $ whose masses
are ${\sqrt{``n^2/R^2"}}$. A rough estimate of the upper limit on $R$ based
on the 100 GeV scale of current experiments and their {1\%} accuracy
yields $R^{-1} \geq 1 TeV $ and this is confirmed by more detailed
{\hbox{comparison${}^{2,3}$}} of the generic predictions of such models
with experiment. The relevant quantum numbers in such theories are
defined by the symmetries of the massless sector and , barring
pathologies of individual models , the enormous phase space
available ensures that any heavy modes initially present will have
long decayed. Conversely their high mass ensures that such modes are
very difficult to excite in the laboratory. Thus the only possibly observable
effects of the heavy excitations come from their participation in
virtual processes and these are also severely suppressed
by their large masses.

Topological solitons (monopoles,vortices,strings.....) are accepted
members of the speculative particle physics zoo. Generically the static  field
configurations $\{\phi_i(x,y)\}$ provide a map from $R^3 X K^d$ into some
configuration space   $ \cal M $ . The
distinct homotopy classes of these maps will then generically include
extremae of the energy and these soliton solutions of the field
equations will manifest as stable particles (modulo quantum effects)
or (if gravity is involved) topologically nontrivial space-time structures.
Such solitons divide naturally into three classes . The special cases

$$\{\phi_i(x)\}  :  R^3 + \{\infty\}\longrightarrow {\cal{M}}\eqno(3a)$$

$$\{\phi_i(y)\}  :  K^d \longrightarrow {\cal{M}}\eqno(3b)$$

correspond respectively to the usual monopoles vortices etc. of four
dimensional physics and the configurations used to provide stress
energy to curve the compactified space $K^d$ in Kaluza-Klein type scenarios.
 The third category

$$\QNAB :\{ R^3 + \{\infty\}\}\quad  X \quad K^d \longrightarrow {\cal{M}}
\eqno(3c)$$

consists of field configurations in which neither the $x$ nor the $y$
dependence of the fields can be contracted away i.e in which the
fields  {\it{ co-wind}} (wind around both spaces simultaneously).
Due to this co-winding attribute we call such configurations {\it{syncyclons}}
(from the greek $\sigma\upsilon\nu$ (with)
$\kappa\upsilon\kappa\lambda o\varsigma$ (turning)). From the point
of view of 4-dimensional  low-energy physics the solitons of type (3.c)  will
appear at distances $>>R$ as stable particles with masses $\sim R^{-1}$
which may possibly have dressed themselves by binding , in an early
epoch,  with lighter particles. Although the Gross-Perry-Sorkin
{\hbox{monopole ${}^{3,4}$}}
can be considered as an example of type (3.c) , since it involves the
extra assumption that the massless U(1) field to which the monopole
couples comes from the higher dimensional graviton multiplet, its
claim to providing a {\it{generic}} signature of extra dimensions is
much weakened for there are many models with gauge fields from other sources.
Moreover the mass of such monopoles is necessarily of the order of the
Planck mass since the radius of the compact space is fixed to be $O(L_p)$
once the gauge coupling is taken to be $O(10^{-2} - 10^{-1} )$. Nevertheless,
inasmuch as the TAUB-NUT solutions of the Euclidean Einstein equations on
$R^3 X S^1$ correspond to gravitational periodic instantons, they are
closely related to the example we shall discuss below which is based
on the periodic instanton or caloron solutions of YM {\hbox{theory ${}^5$}}.

The crucial point is that the topology of maps from a product of
spaces into a compact space is in general much more complex than that
of maps from either space individually. For example although $\pi_1(S^3)=
\pi_2(S^3)=0 $, $S^3$ is well known to be an $S^1$ bundle over $S^2$ (the
Hopf-fibration). Another well known example is furnished by $S^3,S^4$
and $S^7 $.

Yang-Mills theory in 5-dimensions with signature (-++++) is a simple
theoretical studio wherein the basic features of the syncyclonic
scenario come into focus . As we shall see the properties of such
solitons are quite rich and  one seems to be
able to go quite far in predicting the kind of signals that the
non-Abelian sector of the standard model would give if there was an
extra circular dimension.

Thus consider the SU(N) Yang-Mills action on $M^4XS^1$ :

$$   S = -{(\QNK)}\int d^4 x \int_0^R dx^5  {{1}\over 4} {F^a_{MN}}^2
\eqno(4)$$

where $F^a_{MN}= {\partial}_{[M}A_{N]}^a + f^{abc} A^b_M A^c_N,(M,N=0,1,2,3,5
)$. The fields $A_M$ have dimension 1 and the
normalization ($g$ is the 4-dimensional gauge coupling)
 is is fixed by that of the zero mode sector. R is the
perimeter of the extra compact dimension. If we make the static {\it{ansatz}}
${\partial}_0=A_0=0$ the field equations become ${\cal
D}_{\m}F^a_{\m\n}=0\ (\m,\n = 1,2,3,5)$ . The field $A_{\m}({\vec x},x^5=y^1)$
is    independent of time and periodic in y. The energy density is
 $(i,j=1,2,3)$

$$T_{00} = {\QNK}({{1}\over 4}{F^a_{ij}}^2 + {{1}\over 2}{F^a_{5j}}^2)
\eqno(5)$$

 We wish to find
topologically stable finite energy solutions of the 4-dimensional
Euclidean YM equations on $R^3 X S^1$ . The solutions that we are
looking for are thus nothing but the periodic instantons or
{\hbox{{\it{calorons}}${}^5$}}
studied in the context of YM theories at finite temperature !
. We can therefore simply
carry over the results summarized in Ref.6 and reinterpret them in
the present context.

The toplogical data is coded ${}^6$ in the so-called Polyakov operator

$$\QNG = P exp{\int_0^R dx^5 A_5({\vec x},x^5)}\eqno(6)$$

where P denotes path-ordering with respect to $x^5$. The boundary
conditions on the fields are those demanded by finite energy ($ F_{\mu\n}^2
\lnrt r^{-(3+\e)}$  as $r = \mid{\vec x}\mid \lnrt \infty $)
and periodicity in $x^5$. Under periodic gauge transformations
$U({\vec x},x^5+R)=U({\vec x},x^5)$

   $$\QNG \longrightarrow U({\vec x},0) \QNG U^{-1}({\vec x},0)\eqno(7) $$

The eigenvalues \M of \G are therefore gauge
invariant observables. The finite energy boundary conditions imply
that  \M approaches a direction independent limit \N \
as \T  . Thus
 ${\Omega}^{\infty}({\vec x})=V({\vec x})\lambda^{\infty}V({\vec x})^{-1}$
  \ provides, via $V({\vec x})$, a map from  provides a map from $S^2_{\infty}$
into {\hbox {$G/$}} \O where \O is the subgroup of G which commutes with
\N i.e \P \ where  \Q  are the degeneracies \
of the distinct eigenvalues \R \S. The
topology of \G\   for \T \space is thus classified by $\QNU=\QNV=\QNX$ . The
k integers are nothing but the quantized magnetic charges in the
distinct eigenvalue sectors labelled by $\alpha$:

$$\eqalign{\QNH &= \lim_{R\longrightarrow\infty}{{1}\over {4\p i}}
\int_{S^2_{R}}{\QNEE}\cdot tr (P_{\a} {\vec B})\cr
\sum_{\a =0}^kq_{\a} &= 0}\eqno(8) $$

where $P_{\alpha}$ projects ${\vec B}$  onto the $\a$th eigensubspace of \G .
The remaining topology is labelled by a Pontryagin index $\n$ which
arises as follows :\G   can be rewritten as $\QNBB\QNDD{\QNBB}^{-1}$ where
\DD is continuous and well defined throughout $R^3$ while \BB is continuous
on $R^3/{L}$ where L is a set of line singularities corresponding to the
magnetic charges $\{\QNH\}$ . Then

$$\eqalign{Q &= \n(\om^{-1} {\vec{\partial}} \om) +\sum_{\a}
{ {ln\m_{\a}^{\infty}}\over {2 \p i}} q_{\a}\cr
&= \n(\om^{-1} {\vec{\partial}} \om) + {{1}\over {8\p^2} }
\int \QNEE\cdot  tr (ln(\QNG {\vec B}))\cr
\n({\vec J})& = [Q] = {{1}\over {48\p^{2} }}\int d^3{\vec x} \quad
{\vec J}\cdot ({\vec J} X {\vec J})}\eqno(9)$$

Here $[Q]$ is the integer part of the total topological charge Q.
The total topological data ${\QNGG} =(\{{\QNR},q_{\alpha},m_{\alpha}\},\nu)$
  is
invariant under smooth deformations of fields and in the general case
is considerably richer than that for YM theory on $R^4$. This
topological richness is generic to the syncyclonic scenario.
 For each \GG   one
expects to find a solution to the static field equations since the
energy obeys

$${\cal E} = {\QNK} \int ({{1}\over 4}{F^a}^2) \geq {\QNK} \mid
\int {{1}\over 4}(F^a {\tilde F}^a)\mid = {{8\p^2 \mid Q\mid}\over
{g^2 R}} \eqno(10)$$

For the case $\nu=1, q_{\a}=0$   explicit solutions${}^5$
 can be constructed by using the 't Hooft's multi-instanton
 {\hbox{solution${}^7$}} on $R^4$ .
  We specialize the group to SU(2) which is a subgroup
of every SU(N) group.
 In the 't Hooft solution one sets
${\QNJJ}^a=-\QNII{\partial}_{\n}ln {\Pi}$ where \II \ are the anti-selfdual
't Hooft symbols  and $\Pi$ is a scalar function called the superpotential.
The self-duality conditions $\QNE=\QNEB$ then reduce to the linear
equation    ${{\Pi}^{-1}\partial}^2  \Pi =0$ whose solution

$$\Pi(x^{\m}) = 1 + \sum_{n=1}^{K} {{\r_n^2}\over {(x - z_n)^2}}\eqno(11)$$

 describes K instantons with positions $\{z_n\}$ and scales $\{\r_n\}$ in a
singular gauge. To obtain the periodic instanton one
sets $\r_n=\r,z_n=nR\quad {\hat x}^5, n\in{\cal Z}$
  and sums over $n$ ,thus effectively compactifying $x^5$ , and gets${}^5$

$$\eqalign{\Pi({\vec x},x^5)& = 1 + {{\p \r^2}\over {r R}}
 {{\sinh \QNKK}\over {\cosh \QNKK - \cos \QNLL}}\cr
 \QNKK &= {{2 \p r}\over R}\qquad\qquad \QNLL = {{2 \p x^5}\over R}}
 \eqno(12)$$

The gauge potential has a singularity at $r=x^5=0$ which , however, can
be removed by a periodic gauge transformation for instance

$$U= (\sin^2 x^5 + r^2 \cos^2 x^5 )^{-{1\over 2}}(\sin x^5 + i x^j
\cos x^5 \sigma^j)\eqno(13)$$

 The winding number for such configurations (in the physical strip
$x^5\in[0,R]$ ) is 1 and labels the homotopy class of the maps
``$S^2_{\infty}$" $X \quad S^1\longrightarrow S^3$
 furnished by them .We have written ``$S^2_{\infty}$" since in a
singular gauge the winding is actually around an infinitesimal
{\hbox{sphere${}^8$}} centered at $r=x^5=0$ .
 Since ${F^a_{\mu 0}} =0$  they have zero electric field and
thus zero nonabelian charge.Their energy(mass) is $8\pi^2/(g^2R)$.
The scale parameter $\rho$ is not fixed by
the compactification scale R and remains arbitrary. Multi-centered
solutions can be obtained trivially by summing over several distinct
chains separately to obtain :

$$\eqalign{\Pi({\vec x},x^5)& = 1 +\sum_{k=1}^N{{\p {\r}_k^2}\over {r_k R}}
 {{\sinh {\QNKK}_k}\over {\cosh {\QNKK}_k - \cos {\QNLL}_k}}\cr
 {\QNKK}_k &= {{2 \p r_k}\over R}\qquad\qquad {\QNLL}_k = {{2 \p (x^5-x^5_k)}
 \over R}\cr
 r_k &= {\sqrt {{({\vec x} - {\vec x}_k})^2 }}}\eqno (14)$$

and $\{{\vec x}_k,x^5_k\} $ are the positions of the N non-interacting
solitons with total energy $8\pi^2N/(g^2R)$.

In the Kaluza-Klein limit $r>>R$ the fields become

$${ A^a_5\lnrt {{- x^a}\over {r^2( 1 + {\QNOO)}}}} \eqno(15)$$

 $$B_i^a = {1\over 2}\e_{ijk} F^a_{jk} \lnrt -{1 \over {r^2{(1+\QNOO)}^2}}
 [{{x^i x^a}\over {r^2}} - {\QNOO}{ \QNPP}] $$

If $r>> { \QNQQ} $ then one has the fields of a non-Abelian magnetic
dipole (the magnetic moment is $\mu_i^a=({\pi}{\rho}^2/R)\delta_i^a$)
together with a scalar potential :

  $$B_i^a  \lnrt {{\p \r^2}\over R} \QNPP {1\over {r^3}} $$

$$ A^a_5\lnrt -{{\p \r^2}\over R}  {{ x^a}\over {r^3}} \eqno(16)$$

while in the intermediate regime ${\QNQQ} >> r >> R$ the fields are

$$ A^a_5\lnrt - {{ x^a}\over {r^2}} \eqno(17)$$

  $$B_i^a  \lnrt -{ {x^i x^a}\over {r^4}} $$

The energy density $(1/4) {\QNE}^2$ is confined${}^6$ to the smallest scale
available i.e $\r$ if $\r\approx R$ or $\r<<R$ and R if $\r>>R$. Thus
classically such solitons will appear as pointlike particles with
masses O(\K) and dipolar asymptotic fields which become monopolar at
short distances ($>>R$\ !).

The gauge field posesses zero modes associated with translations (translation
$=$ rotation on the internal circle !) dilatations and global gauge
rotations. Collective coordinates must be introduced for each of
these and quantized by the standard semiclassical methods ${}^8$
to find the quantum states of the soliton    . Moreover since we
expect the the YM fields to be coupled to light fermions in any realistic
example fermionic zero modes should also be taken into account. This
leads to fermion number fractionalization and the induction of
non-Abelian charges on the {\hbox{soliton ${}^{12}$}}. Barring
{\hbox{pathologies${}^9$}},the quantization of the
collective coordinates associated with global gauge rotations implies
that the states of the soliton occupy representations of the YM
group. It is interesting to note that since the asymptotic falloff of
the fields in the present case is dipolar the
{\hbox{``breakdown of color''${}^{9,10}$}}
 that takes place when the corresponding
collective coordinates of non-Abelian (grand unified) 't
Hooft-Polyakov monopoles are quantized should be evaded
{\hbox{since${}^{10}$}} this breakdown has been shown to be connected
with the $r^{-2}$ falloff of the color magnetic fields and is absent
in the case of a monopole-antimonopole system with $r^{-3}$
asymptotics. These questions require detailed analysis presently in
progress in collaboration with V.Soni and will be reported on subsequently.

A preliminary sketch of the situation is as follows. In keeping with
our basic aim of trying to deduce model independent consequences of
the existence of extra compact dimensions let us take the YM group to
be SU(3) color in which the SU(2) group associated with the soliton
is embedded trivially. In that case it is reasonable to assume that
the scale parameter $\r$ is smaller than $1 {GeV}^{-1}$ since at such
scales QCD becomes confining and classical considerations are largely
irrelevant. On scales $R<<r<<1{Gev}^{-1}$ the QCD coupling is weak
and semi-classical heuristics are applicable. The fermion masses of
the standard model with \J \ flavors are essentially all negligible
compared to \K$> 1TeV$ and can at most lead to small splittings of
the degenerate ground states that arise due to fermion number
fractionalization. \J \  fermionic zeromodes {\hbox{develop ${}^{11}$}}
 and for antiperiodic boundary conditions around $S^1$ their wave
  functions are exponentially localized ($e^{-r/R})$
around the soliton. A degeneracy of ground states associated with the
zeromode being occupied or not develops. A baryon number

$$\eqalign{\langle B \rangle &= {1\over 3}\langle \int d^4 x
\sum_{i=1}^{N_f} \QNTT \QNSS \rangle\cr
&=-{{N_f}\over {192\p^2}}\int d^4 x\quad  tr(F\tilde F)\cr
&= -{{N_f}\over 6} Q}\eqno(18)$$

is induced . Similarly one gets $B/2$ for the
electric charge. Using the result of Ref.12 for the
induced color charge one finds

$$\eqalign{\langle Y^A \rangle &= \langle \int d^4 x
\sum_{i=1}^{N_f} \QNTT Y^A \QNSS \rangle\cr
&=-{{{}N_f}\over {64\p^2}}\int d^4 x\quad  tr(Y^A F\tilde F)\cr
&= -{{N_f}\over 2} Q \d^{A8}}\eqno(19) $$

$(A,B=1,2...8)$ and the quark hypercharges have been normalized to (1,1,-2).
The effective action ${{\Gamma}_{ind}}$ from which this expectation value may
be deduced as $\int{{{\d{{\Gamma}_{ind}}}\over{\d {A_{0}}}}}$ is the
Chern-Simons
term in 5 dimensions ${}^{12}$. An analogy with the Skyrmeon problem
${}^{13}$ with the roles of flavor and color interchanged is evident and
a six-dimensional representation (i.e ${\int_{M^6} tr F^3}$
with $\partial M^6=M^4 X S^1$ ) of the induced
action is available to describe the topologically non-trivial
sectors of such theories.

Since the ground states pick up a color charge the soliton should now
have a nonzero color electric field and this will result in the
formation of a bag at scales $O(1 {GeV}^{-1})$ with quarks trapped
within it to neutralize the color charge of the semiclassical states
discussed above. Similar scenarios may be spelt out for the
Salam-Weinberg model at scales $ r<<(10^2 GeV)^{-1}$ where the v.e.v of
the scalar field is negligible and will result in lepton number
carrying objects with large masses. It is interesting to note that the map
provided by the Standard Model doublets from ${S^2}_{\infty} X S^1$ to $S^3$
can be noncontractible in contrast to the usual situation.

Let us now briefly sketch how stringlike and sheetlike solitonic
structures  could arise . Consider a YM theory in 6 dimensions of which
two are circular (Self-Dual Euclidean solutions on $R^2 X S^2$ could also be
used).
Let \VV\ be coordinates for $M^4$ and $(x^5=y^1,x^6=y^2)$
those for the extra dimensions of radii $(R_1,R_2)$. Then for field
configurations $A_{\mu}$\VV,\XX \ independent of time and $x^3$ the field
equations are as before except that now one must solve them on \YY.
The same proceedure used for going from solutions on $R^4$ to
soloution on \D may now be used to go from \D\ to \YY since the
multisoliton solutions (14) are available for arbitrary locations
of arbitrary numbers of solitons. Thus the superpotential for our
stringlike objects is

$$\eqalign{\Pi({\vec x},y_1,y_2)
&=1+ \sum_{(n_1,n_2) \in {\cal Z}^2} {{\r^2}\over {
{\vec x}^2 + (y_1 - n_1 R_1)^2 + (y_2 - n_2 R_2)^2}}\cr
&= 1 +\sum_{n\in {\cal Z}} {{\p \r^2}\over {r_n R_1}}
 {{\sinh \QNKK_n}\over {\cosh \QNKK_n - \cos \QNBBB}}\cr
\QNKK_n &={ {2\p} \over R_1} {\sqrt{x_1^2 +x_2^2 + (y_2 -n R_2)^2}}=
{ {2\p} \over R_1}r_n\cr
 \QNBBB&= { {2\p} \over R_1}y_1}\eqno (20)$$

It is obviously  periodic in $(y^1,y^2)$ . Actually since the series
above has a divergence the above formula needs a subtraction
proportional to $\zeta(1)$ ,after which it displays the logarithmic
asymptotics in $\r={\sqrt{x_1^2+x_2^2}}$ expected for solutions of
the equation for the superpotential in two non-compact
dimensions. The winding number {\it{per unit length}} in the physical
strip $y^i\in [0,R_i)$ is obviously again one and is assosciated with
the homotopy of maps \ZZ. Thus the energy per unit length  is O(\K).
Once again coupling to fermions will induce fermion number,charge and
non-Abelian charge on the string. A cylinderical bag  may be expected to form
around QCD strings of this type at distances from the core of order
$1 (GeV)^{-1}$ . The cosmological properties of such structures may provide
a window of allowed values of the compactification scale.

The same proceedure can be repeated to include sheetlike defects.  The
relevant homotopy is that of maps $\QNCCC \lnrt S^3$ ,where the
 $Z_2$ is assosciated with the two sides of the sheet .

To conclude we have utilized the known properties of an available
{\hbox{ solution${}^{5,6}$}} of the Euclidean YM field equations in 4
dimensions
to illustrate the   complexity and
richness of the generic scenario of co-winding solitons
that may well be a common and relatively model independent
 necessary implication of the
plethora of higher dimensional models proposed over the last few years.
This model independence and the topological stability of such solitons
combined with the fact that the {\it{known}} gauge group of the
standard model will be sensitive to the extra dimensions through the
cowinding channel singles out syncyclons as one of the few possible
distinct signatures of higher dimensional scenarios.

{\bf{Acknowledgements :}} I am grateful to G.Senjanovic, S.Ouvry,
 J.McCabe, A.Comtet, Prof. I. Todorov, S. Phatak ,  and
 A.Khare for helpful discussions ,to J.P.T\'ecourt for etymological
 advice, to V.Soni for instructive collaboration and to the Theory
  Group at IPN,Orsay, where this work was largely done, for their warm and
   pressure free hospitality.

{\bf{Bibliographical Note}} : After this idea was reported at a
seminar at ICTP, Trieste we were informed by S.Randjbar-Daemi that Strominger
${}^{14}$ had made some remarks along the same lines. The ``five-branes''
of Ref.14 are also based on instanton solutions. Their behaviour
under compactification of some of the coordinates is interesting.

\vfil

\centerline {\bf{References }}

\item {1)} A.Salam and J.Strathdee, Ann. of Physics 141 (1982) 316.
\item {2)} V.A.Kosteleck\'y and S.Samuel, Phys. Lett. B270 (1991)21.
\item {3)} R.Sorkin, Phys. Rev. Lett. 51 (1983) 87.
\item {4)} D.Gross and M.Perry, Nucl. Phys. B226 (1983) 29.

\item {5)} B.J.Harrington and H.K.Shepard, Phys. Rev. D17 (1978), 2122.
\item {6)} D.Gross, R.Pisarski, L.Yaffe, Rev. Mod. Phys. 53 (1981) 43.
\item {7)} G.'t Hooft (unpublished);Phys. Rev. D14(1976)3432.
\item {} E.Corrigan and D.Fairlie, Phys. Lett. B67 (1977), 69.
\item {} R.Jackiw, C.Nohl. and C.Rebbi, Phys. Rev. D15, (1977)1642.
\item {8)} R.Rajaraman, {\it Solitons and Instantons}, North Holland,
1982 and references therein.
\item {9)} A.Abouelsaood, Nucl. Phys. B226 (1983) 309.
\item {} P.Nelson, Phys. Rev. Lett. 50 (1983), 939.
\item {} P.Nelson and A.Manohar, Phys. Rev. Lett. 50 (1983) 943.
\item {} C.P.Dokos and T.N.Tomaras, Phys. Rev. D 21 (1980), 2940.
\item {10)} S.Soleman and P.Nelson, Nucl. Phys. B237 (1984)1.
\item {11)} R.Jackiw and C.Rebbi, Phy. Rev. D13 (1976) 3398.
\item {} B.Grossman, Phys. Lett. A61 (1977), 86.
\item {12)} A.Niemi and G.W.Semenoff, Physics Reports 135 (1986) 99;
\item{}Phys. Rev. Lett. 51 (1983) 2077.
\item {13)} E.Witten, Nucl. Phys. B223 (1983) 422.
\item{} S.Jain and S.R.Wadia , Nucl. Phys. B258(1985) 713.
\item {14)} A.Strominger Nucl. Phys. B 343 (1990) 167.
\vfil
\eject
\end